\documentclass[twocolumn,secnumarabic,amssymb, nobibnotes, aps, prd]{revtex4-1}

\usepackage{amssymb,amsmath,amsfonts}
\usepackage{bbm,wasysym}
\usepackage{graphicx}
\usepackage{xcolor}
\usepackage{makecell}
\usepackage{times}
\usepackage{verbatim}
\newcommand{\id}{{\mathbbm{1}}}

\newcommand{\bra}[1]{{\langle #1 |}}
\newcommand{\ket}[1]{{| #1 \rangle}}

\newcommand{\tr}{tr}

\begin{document}

\title{A review on super dense coding over covariant noisy channels}%
\author{Z. Shadman}
\email[]{cqtzs@nus.edu.sg}
\affiliation{ Centre for Quantum Technologies, National University of Singapore, 3 Science Drive 2,  117543 Singapore, Singapore}
\author{ H. Kampermann}
\author{D. Bru\ss}%
\affiliation{ Institute f\"ur Theoretische Physik III,
              Heinrich-Heine-Universit\"at
              D\"usseldorf, D-40225 D\"usseldorf, Germany}
\author{ C. Macchiavello}
\affiliation{Dipartimento di Fisica and INFM-Unit$\acute{a}$ di Pavia, Via Bassi 6, 27100  Pavia, Italy}

\begin{abstract}

We study the performance of the super dense coding protocol in the presence
of quantum channels with covariant noise. We first consider the bipartite case
and review in a unified way the case of general Pauli channels.
We discuss both the cases of unitary and non-unitary encoding.
We also study the multipartite scenario and investigate  the case of many
senders and one receiver.

\end{abstract}


\maketitle

\section{Introduction}

The super dense coding protocol was originally proposed in Ref. \cite{Bennett}
and exploits the non-local properties of entangled states in order to
communicate two bits of classical information by sending one
qubit only. After the original proposal, the super dense
 coding protocol was studied in various scenarios over noiseless channels
and with unitary encoding \cite{hiroshima, ourPRL, Dagmar}.
Super dense coding in the presence
of noise along the transmission channels was then investigated
in \cite{zahra-paper} for uncorrelated noise (i.e. memoryless channels)
and in \cite{zahra-mem} for the case of correlated noise, i.e.
when the noisy channel acting on two subsystems cannot be expressed as a
product of two independent channels acting on each subsystem separately
(such kinds of channels were originally
analysed from the point of view of optimisation of the classical information
transmission \cite{mp,memory-quasi-chiara,cerf}).

The notion of multipartite super dense coding, namely the case of
more than one sender and/or more than one receiver, was introduced
by Bose \emph{et al.} \cite{Bose-multi-first} to generalise
the original scheme \cite{Bennett} of super dense coding to
multiparties. Distributed super dense coding was  also widely discussed in
\cite{ourPRL,Dagmar}, where two scenarios of many senders with
either one or two receiver(s) were addressed and it was shown that
bound entangled states with respect to a  bipartite cut between the senders
(Alices) and the receiver (Bob) are not ``multi'' dense-codeable.
Moreover, a general classification of multipartite quantum states according
to their dense-codeability was investigated. The case of noisy channels in
the multipartite scenario was studied more recently in \cite{zahra-multi}
for multi-partite covariant channels. In the present work we review in a
unified way the main results achieved for super dense coding over covariant
noisy channels, originally reported in
\cite{zahra-paper,zahra-mem,zahra-multi}.

The paper is organised as follows. In section \ref{1-1-0} we review
the protocol and the Holevo bound for the super dense
coding capacity in the presence of an arbitrary channel $\Lambda$.
Section \ref{1-3} is devoted to the super dense coding
capacity in the presence of bipartite covariant channels and both unitary and  non-unitary  encoding. In section \ref{s:pauli}, we study the case of Pauli channels
as an example for covariant channels.
In section \ref{s:nu},  for some explicit examples, we discuss the advantage of  non-unitary encoding over unitary encoding.
In section \ref{s:multi} we consider the multi-partite scenario.
Finally, in section \ref{conc}, we summarise the main results.
\section{The protocol and its capacity \label{1-1-0}}

In the original (noiseless) super dense coding protocol
one starts from a $d^2$-dimensional bipartite shared state $\rho$
between the sender Alice and the receiver Bob.
Alice performs with probability $p_i$ a local
   unitary operation $W_i$  on her subsystem to encode  classical
information through the state
$\rho_i= (W_i\otimes\id)\rho({W_i}^{\dagger}\otimes\id).$ Subsequently,
she sends her subsystem to Bob
(ideally via a noiseless channel). The
ensemble that Bob receives is $\{\rho_i,p_i\}$.

The performance of a given composite state $\rho$ for super dense coding is
usually quantified by the Holevo quantity, maximised over all possible
encodings on Alice's side.
A theorem stated by Gordon \cite {Gordon} and Levitin  \cite{Levitin},
and proved by Holevo \cite {Holevo-chi-quantity}, states
that the amount of accessible classical information ($I_{acc}$)
contained in an ensemble $\{ \rho_i, p_i\}$ is upper bounded by
the so-called $\chi$-quantity $\chi(\{ \rho_i, p_i\})$, often also referred to as Holevo quantity. This upper
bound holds for any measurement that can be performed on
the system, and is given by
\begin{eqnarray}
I_{acc}\leq \chi(\{ \rho_i, p_i\})\equiv
S\left(\overline{\rho}\right)-\sum_i p_i S\left(\rho_i\right),
\label{chi-quantity1}
\end{eqnarray}
where $\overline{\rho}=\sum_i p_i \rho_i$ is the average ensemble
state and $ S(\eta)=-\mathrm{\tr}(\eta \log \eta)$ is the von Neumann
entropy of $\eta $. From the concavity of the von Neumann entropy
$S(\rho)$ it follows that the Holevo quantity is non-negative.
The Holevo bound (eq. \ref{chi-quantity1}) is achievable in
the asymptotic limit \cite{Schumacher-Westmoreland, Holevo-capacity}.

The maximal amount of the Holevo quantity that can reliably be transmitted
in this process with respect to the unitary operators $W_i$,
chosen with the probabilities $p_i$ is then known as super dense coding
capacity and it is given by
\begin{eqnarray}
 C&=&\max_{\{W_i,p_i \}}(\chi{\{\rho_i,p_i\}})\nonumber\\
&\equiv& \max_{\{W_i,p_i \}}
 \left(S\left(\overline{\rho}\right)-\sum_i p_i
S\left(\rho_i\right)\right).
\end{eqnarray}
It has been shown that for noiseless channels and unitary encoding,
the capacity is given by $C=\log d+S(\rho_b)-S(\rho)$ \cite{ziman,hiroshima}.
Here, $\rho_b$ is Bob's
reduced density operator with $\rho_\textmd {b}=\mathrm{tr_a}\rho$.
Without the additional resource of entangled states, a
$d$-dimensional quantum state can be used to transmit the
information  $\log d$. Hence, quantum states  for which
$S(\rho_b)-S(\rho)>0 $, i.e. those which are more mixed locally
than globally, are the useful states for super dense coding.

Let us now consider the noisy case, namely the scenario where the state $\rho$
is transmitted over noisy channels.
A noisy quantum channel can be described in general as a completely positive
trace preserving (CPTP) map acting  on the quantum state that is transmitted.
We consider
$\Lambda:\rho_i \rightarrow \Lambda(\rho_i)$ to be a CPTP map that
acts on the encoded  state $\rho_i= (W_i\otimes\id)\rho({W_i}^{\dagger}\otimes\id)$. For $\{\Lambda(\rho_i,p_i)\}$
being the ensemble that Bob receives, the Holevo quantity is given by
\begin{eqnarray}
\chi\{\Lambda(\rho_i,p_i)\}&=&S\left(\overline{\Lambda(\rho)}\right)-\sum_i
p_i S\left(\Lambda(\rho_i)\right)\nonumber\\
&=&\sum_{i}p_iS\left(\Lambda(\rho_i) \vert \vert \overline{\Lambda(\rho)}\right),
\label{Holevo}
\end{eqnarray}

where $\overline{\Lambda(\rho)}=\sum_i p_i \Lambda(\rho_i)$ is the
average state after transmission through the noisy channel and
$S(\rho \vert \vert \sigma)=\mathrm{\tr} \rho \left(\log
\rho - \log \sigma \right)$ is the relative entropy.
The super dense coding capacity $C$ for a given resource state $\rho$ and the
noisy channel $\Lambda$ then takes the form
\begin{eqnarray}
 C&=&\max_{\{W_i,p_i \}}(\chi{\{\Lambda(p_i,\rho_i)\}})\nonumber\\
&\equiv& \max_{\{W_i,p_i \}}
 \left(S\left(\overline{\Lambda(\rho)}\right)-\sum_i p_i S\left(\Lambda(\rho_i)\right)\right).
\end{eqnarray}
In the following, we will concentrate on the optimisation
of the Holevo quantity
in order to find the super dense coding capacity
for covariant noisy channels  and unitary encoding as well as non-unitary encoding.

\section{Super dense coding with general encoding over covariant noisy channels\label{1-3}}

{For noiseless channels, the super dense coding protocol with non-unitary encoding
 has been discussed by
M. Horodecki and Piani \cite{CPTP}, M. Horodecki et al. \cite{unitaryoptimal1},
and Winter \cite{unitaryoptimal2}. In this section we consider the possibility of performing non-unitary encoding in the presence of noisy channels    that are covariant for a
complete set of orthogonal unitary operators
$\tilde{V}_{i}$, namely these channels have the property

\begin{eqnarray}
\Lambda^\textmd{c}_{ab}(\tilde{V}_{i}
\rho   \tilde{V}_{i}^\dagger) =
  \tilde{V}_{i} \Lambda^\textmd{c}_{ab}(\rho)
\tilde{V}_{i}^\dagger,
\label{covariance1}
\end{eqnarray}
for the set of unitary operators which satisfy the orthogonality
condition ${\mathrm{\tr}}[\tilde{V}_{i}\tilde{V}_{j}^\dagger]=d \delta_{ij}$.
According to \cite{hiroshima}, for this set it is guaranteed that $ \frac{1}{d}\sum_i \tilde{V}_{i} \Xi {\tilde{V}_{i}^\dagger} = \id \mathrm{\mathrm{\tr}} \Xi$  where $ \Xi$ is an arbitrary, $d$-dimensional, operator.   In eq. (\ref{covariance1}), $\Lambda_{ab}^c$ is a CPTP map which indicates a covariant channel. Here we are interested in unitary operators $\tilde{V}_{i}$ of the form
\begin{eqnarray}
\tilde{V}_{i}= \tilde{V}_{i}^{a}\otimes  \id^{b}.
\label{Vtilde2}
\end{eqnarray}

{For the super dense coding protocol with non-unitary encoding and in the presence of a covariant channel,} let us consider $\Gamma_i$ to be a completely positive trace preserving
(CPTP) map. Alice applies the map $\Gamma_i$ on her side of the shared state
$\rho$, thereby encoding $\rho$ as $ \rho_i= [\Gamma_i \otimes \id]
 (\rho):=\Gamma_i(\rho)$. The rest of the scheme is similar to the
 case of unitary encoding. Alice sends the encoded state
 $ \rho_i=\Gamma_i(\rho)$ with the probability $p_i$ to Bob
 through the covariant channel $\Lambda_{\textmd {ab}}^c $.
 Now, we are interested in understanding which ensemble of CPTP maps achieves
 the super dense coding capacity and what is the optimum  Holevo quantity
with respect to the encoding $\Gamma_i$ and $p_i$.
The optimisation of the Holevo quantity  is given in the  following lemma. \\

\textbf{Lemma 1.} Let
\begin{eqnarray}
\chi_ {\textmd {non-un}}&=&S\Big (
\sum_{i} p_{i}
 \Lambda_{\textmd {ab}}^\textmd {c}\left(\Gamma_i(\rho) \right)\Big)\nonumber\\
&-&\sum_{i} p_{i}
S\Big(\Lambda_{\textmd {ab}}^\textmd {c} \left(\Gamma_i(\rho)\right)\Big),
 \label{Holevo-co}
\end{eqnarray}
be the Holevo quantity with non-unitary encoding  $\Gamma_i(\rho):= [\Gamma_i \otimes \id]
 (\rho)$ and $\Lambda_{\textmd {ab}}^\textmd {c}$  be a
covariant channel with the property
(\ref{covariance1}).
Let $\Gamma_ {\textmd{min}}(\rho):=[\Gamma_{\textmd{min}} \otimes \id](\rho)$
 be the map that minimises the von Neumann entropy  after application of
this map and  the covariant channel $\Lambda^{\textmd c}_{\textmd {ab}}$
to the initial state $\rho$, i.e. $\Gamma_{\textmd{min}}$ minimises the
expression $S\left(\Lambda^{\textmd c}_{\textmd {ab}}(\Gamma_{\textmd{min}}
(\rho))\right)$.
 Then the super dense coding capacity with non-unitary encoding  $C_{\textmd{non-un}}$ is given by

\begin{eqnarray}
 C_{\textmd {non-un}}=\log d + S\Big(\Lambda^\textmd {}_\textmd{b} (\rho_\textmd{b})\Big)-
    S\Big(\Lambda^{\textmd c}_{\textmd {ab}}\big(\Gamma_{\textmd{min}}(\rho)\big)\Big).
\label{capacity-non-un}
\end{eqnarray}

where  $d_\textmd{}$ is the dimension of Alice's Hilbert space,  and
$\mathrm{\tr}_{\textmd{a}}  \Lambda_{\textmd{ab}}^\textmd{c}
\left(\rho^ {\textmd{}} \right)=\Lambda_{\mathrm{b}}^\textmd{}
\left(\rho_{\mathrm{b}}\right)$. \\

\textbf{Proof:} The von Neumann entropy is sub-additive. The maximum entropy of
a $d$-dimensional system is $\log d $.
Since  $\Gamma_{\textmd{min}}$  is a map that leads to the minimum
of the output  von Neumann entropy, an upper bound on Holevo quantity
(\ref{Holevo-co})  can be given  as
\begin{eqnarray}
\chi_ {\textmd {non-un}}&\leq& S\Big(\sum_{i} p_{i}
 \Lambda_{\textmd{ab}}^\textmd{c} \left(\rho_{i}  \right)\Big)- S\Big(\Lambda^{\textmd c}_{\textmd {ab}}\big(\Gamma_{\textmd{min}}(\rho)\big)\Big)\nonumber\\
&\leq& \log d_\textmd{}+ S\left(\Lambda_{\textmd{b}}^\textmd{}
(\rho_{\textmd{b}})\right) -    S\Big(\Lambda^{\textmd c}_{\textmd {ab}}\big(\Gamma_{\textmd{min}}(\rho)\big)\Big).
\label{upper-bound-multi}
\end{eqnarray}
In the next step, we show that the upper bound (\ref{upper-bound-multi}) is
reachable by the ensemble  $\{{\tilde{\Gamma}_i(\rho)},\tilde{
p}_i\}$ with $\tilde{
p}_i=\frac{1}{d^2}$ and
$\tilde{\Gamma}_i(\rho)=(\tilde{V}_{i}  \otimes\id) \Gamma_{\textmd{min}} (\rho)(\tilde{V}_{i} ^\dagger \otimes\id)$ where $\tilde{V}_{i} $ was
defined in eqs. (\ref{covariance1}) and (\ref{Vtilde2}).

 The Holevo quantity for
the ensemble  $\{{\tilde{\Gamma}_i(\rho)},\tilde{
p}_i\}$  is denoted by
$\tilde{\chi}_{\textmd {non-un}}$ and is given by

\begin{eqnarray}
\tilde{\chi}_{\textmd {non-un}}&=&S \Big( \sum_i\frac{1}{d^2}\Lambda^{\textmd c}_{\textmd {ab}}\left(\tilde{\Gamma}_i(\rho)\right)\Big)\nonumber\\
&-&\sum_i \frac{1}{d^2}  S\left(\Lambda^{\textmd c}_{\textmd {ab}}\left(\tilde{\Gamma}_i(\rho)\right)\right).
\label{optiensemble-non-un}
\end{eqnarray}

By using the covariance property (\ref{covariance1}), the argument in
the first term on the RHS of (\ref{optiensemble-non-un}) is given
by
\begin{eqnarray}
&&  \sum_i\frac{1}{d^2}\Lambda^{\textmd c}_{\textmd {ab}}\left(\tilde{\Gamma}_i(\rho)\right)\nonumber\\
 &=& \frac{1}{d^2} \sum_{{i}} (\tilde{V}_{{i}}\otimes
\id^\textmd{b} )
\Lambda_{\textmd{ab}}^\textmd{c} \big(\Gamma_{\textmd{min}} (\rho)  \big)
(\tilde{V}_{{i}}^\dagger\otimes \id^\textmd{b})\nonumber\\
&=& \frac{1}{d^2} \sum_{{i}} (\tilde{V}_{{i}}\otimes
\id^\textmd{b} ) \varrho(\tilde{V}_{{i}}^\dagger\otimes \id^\textmd{b}),
\end{eqnarray}
where $\varrho:=\Lambda_{\textmd{ab}}^\textmd{c} \big(\Gamma_{\textmd{min}} (\rho)  \big)$. The density matrix  $\varrho$
in the Hilbert-Schmidt representation can be  decomposed as
\begin{eqnarray}
\varrho &=&\frac{\id^{\textmd{}}}{d}\otimes\Lambda _{\textmd{b}}
(\rho_\textmd{b})+\sum_j r_j\lambda_j^{\textmd{}}
\otimes\id^{\textmd{}}
+\sum_{j,k}t_{jk} \lambda_j^{\textmd{}}\otimes\lambda_k^\textmd{}\;,\nonumber\\
\label{decomposition-multi}
\end{eqnarray}
where the  $\lambda_j^{\textmd{}}$ are the generators of the	
$SU(d)$ algebra with $\textmd{\tr}\lambda_j=0$.
The parameters $r_j$ and $t_{jk}$ are real numbers.
By exploiting  the equation
$ \frac{1}{d}\sum_i \tilde{V}_{i} \Xi \tilde{V}_{i} ^\dagger =
\id \mathrm{\mathrm{\tr}} \Xi$, and since each $\lambda_j$ is traceless,
we can write
\begin{eqnarray}
 \sum_{i}
\tilde{V}_{i} \lambda_j^{\textmd{}}  {\tilde{V}}_{i}^{\dagger}=0.
\label{average-lambda}
\end{eqnarray}
By using this property and the decomposition (\ref{decomposition-multi}),
we find that the argument in
the first term on the RHS of (\ref{optiensemble-non-un}) is given
by

\begin{eqnarray}
 S \bigg( \sum_i\frac{1}{d^2}\Lambda^{\textmd c}_{\textmd {ab}}\left(\tilde{\Gamma}_i(\rho)\right)\bigg)
=\log d+ S\left(\Lambda_{\textmd{b}}^\textmd{} \left(
\rho_{\textmd{b}}\right)\right).
\label{average-state1}
\end{eqnarray}

Furthermore, the second term on the RHS of eq. (\ref{optiensemble-non-un})
can be expressed in terms of the
unitary operator $\Gamma_{\textmd{min}}$ and the channel. By using the covariance property (\ref{covariance1}),  and since the von Neumann entropy is invariant under a unitary
transformation, we can write
\begin{eqnarray}
&&\sum_i \frac{1}{d^2}  S\left(\Lambda^{\textmd c}_{\textmd {ab}}\left(\tilde{\Gamma}_i(\rho)\right)\right).\nonumber\\
&=&\frac{1}{d^2} \sum_{{i}} S\bigg((\tilde{V}_{{i}}\otimes
\id^\textmd{b} )
\Lambda_{\textmd{ab}}^\textmd{c} \big( \Gamma_{\textmd{min}} (\rho) \big )
(\tilde{V}_{{i}}^\dagger\otimes \id^\textmd{b})\bigg)\nonumber\\
&=& S\Big(\Lambda^{\textmd c}_{\textmd {ab}}\big(\Gamma_{\textmd{min}}(\rho)\big)\Big).
\label{average-entropy1}
\end{eqnarray}

 Inserting eqs. (\ref{average-state1}) and (\ref{average-entropy1}) into
eq. (\ref{optiensemble-non-un}), one finds that the Holevo quantity
$\tilde{\chi}_{\textmd {non-un}} $ is equal to  the upper bound
given in
eq. (\ref{upper-bound-multi}) and therefore this is the super dense
coding capacity. \hfill {$\Box$}

As we can see from the capacity expression (\ref{capacity-non-un}),
all the parameters are known except the single CPTP map
$\Gamma_{\textmd{min}}$.

{  When we restrict our super dense coding protocol to unitary encodings, the capacity given in eq. (\ref{capacity-non-un})  changes to the following 
expression

\begin{eqnarray}
C_{\textmd{un}}&=&\log d+ S\left(\Lambda_{\textmd{b}}^\textmd{}
\left(\rho_{\textmd{b}}\right)\right)\nonumber\\
&-& S\bigg(\Lambda_{\textmd{ab}}^\textmd{c} \Big(\big(U_{\textmd{min}}
 \otimes \id^\textmd{b} \big)  \rho^{\textmd{}} \hspace{1mm}
\big(U_{\textmd{min}}^{\dagger}\otimes\id^\textmd{b} \big)\Big)\bigg),
\label{dc-covariant}
\end{eqnarray}
where, similarly to $\Gamma_{\textmd{min}}$,  the unitary operator   $U_{\textmd{min}}$  minimises the von Neumann entropy after application of this
unitary operator and  the channel  $\Lambda_{\textmd {ab}}^\textmd {c}$  to
the initial state $\rho^{\textmd {}} $. For some specific situations like  noiseless channels,
i.e. for $ \Lambda^{\textmd c}_{\textmd{ab}}=\id$,
 this unitary operator  $U_{\textmd{min}}$ has  already been identified as the identity operator.
We also provide more examples in the next section. }}

\section{Pauli channels\label{s:pauli}}

In this section we will address the case of Pauli channels as an example
of covariant channels described above. Consider first a Pauli channel acting
on a single quantum system  described by a $d$-dimensional density
operator $\xi$. The action of the Pauli channel can be written as
\begin{eqnarray}
\Lambda^{\textmd {P}}(\xi)=\sum_{m,n=0}^{d-1}q_{mn} V_{mn}\xi V_{mn}^\dagger\;,
\label{pauli-d-channel}
\end{eqnarray}
where $q_{mn}$ are probabilities (i.e. $q_{mn}\geq 0$ and
$\sum_{mn}q_{mn}=1$). The unitary \emph{displacement} operators $V_{mn}$ are
defined as
\begin{eqnarray}
V_{mn}=\sum_{k=0}^{d-1}\exp \left({\frac{2i\pi kn}{d}}\right)\ket{k}\bra{k+m(\mathrm{mod}\,
{d})}\;.
\label{vmn}
\end{eqnarray}
The above operators satisfy $\mathrm{\tr} V_{mn} = d\delta_{m0} \delta_{n0} $
and $V_{mn} V_{mn}^\dagger=\id$, and commute up to a phase,
\begin{eqnarray}
 V_{mn} V_{\tilde{m}\tilde{n}}
=\exp\left({\frac{2i\pi(\tilde{n}m-n\tilde{m})}{d}}\right) V_{\tilde{m}\tilde{n}}V_{mn}.
\label{vv}
\end{eqnarray}
As the  operators $V_{mn}$ in  eq. (\ref{pauli-d-channel}) are unitary, the
Pauli channel is unital, i.e. it preserves the identity. For a bipartite
system a general Pauli channel can be defined as
\begin{eqnarray}
\Lambda^{\textmd P}_{\textmd {ab}}(\xi)=\sum_{m,n,\tilde{m},\tilde{n}=0}^{d-1}
 q_{mn\tilde{m}\tilde{n}} (V_{mn}\otimes V_{\tilde{m}\tilde{n}})\xi
  (V_{mn}^\dagger\otimes V_{\tilde{m}\tilde{n}}^\dagger),\nonumber\\
\label{2-Pauli-channels}
\end{eqnarray}
 where $q_{mn\tilde{m}\tilde{n}}$ are probabilities. The above form includes
several physical situations, such as for example the case of a 
one-sided channel,
where noise acts only on one subsystem (during the transmission of Alice's subsystem to Bob), or a two-sided channel, where noise acts on both subsystems
$A$ and $B$.
In the latter case noise can be independent on the two subsystems, namely
the probability distribution $q_{mn\tilde{m}\tilde{n}}$ is factorised as
$q_{mn\tilde{m}\tilde{n}}=q_{mn}q_{\tilde{m}\tilde{n}}$, or it may exhibit
correlations and the probability does not have the simple factorised form
\cite{mp,memory-quasi-chiara,cerf}. In the following we will consider in
particular the family of correlated channels modelled as
$q_{mn\tilde{m}\tilde{n}}=(1-\mu)q_{mn}q_{\tilde{m}\tilde{n}}
+\mu q_{mn}\delta_{m,\tilde{m}}\delta_{n,\tilde{n}}$, where the parameter
$\mu$ ($0\leq\mu\leq1$) quantifies  the correlation degree. For the particular
value
 $\mu=0$ the two channels $\Lambda^P_a$ and $\Lambda^P_\textmd {b}$ are
uncorrelated and act independently on Alice's and Bob's subsystems,
respectively. For $\mu=1$ the channel
is called \emph{fully
  correlated} and for other values of $\mu$, different from
zero and one, the channel is partially correlated.
In the next sections we will study the above different scenarios.

\subsection{One-sided channel \label{one}}

In the one-sided channel noise acts only on the subsystem that is transmitted
by Alice to Bob after encoding and it is a particular case of eq.
(\ref{2-Pauli-channels}) with
$V_{\tilde{m}\tilde{n}}=I$. This case was extensively studied in Ref.
\cite{zahra-paper}.
We report here the main results, related to the cases of shared Bell and
Werner states.

A Bell state in $d$ dimensions is defined as
$\ket{\psi_{00}}= \frac{1}{\sqrt d}\sum_{j=0}^{d-1}\ket{j}\otimes \ket{j}$.
The set of the other maximally entangled Bell states is then denoted by
$\ket{\psi_{mn}}=(V_{mn}\otimes\id)\ket{\psi_{00}} $, for $m,n=0,1,...,d-1$.
In this case it can be proved that the output von Neumann entropy is 
independent of the unitary operator $U_{min}$ and it is equal to the Shannon 
entropy $H(\{q_{mn}\})$ \cite{zahra-paper}. We also notice that for a Bell 
state the reduced state is $\rho_b=\frac{\id^{\textmd{}}}{d}$. Therefore,  according to eq. (\ref{dc-covariant}),  the super dense coding capacity  for an input Bell state takes the  form

\begin{eqnarray}
C_{\mathrm{un,B}}^{\mathrm{P,  one-sided}}
&=&\log d^2-H(\{q_{mn}\})
\label{DCcapacityd-qubitcha}
\end{eqnarray}
where $H(\{q_{mn}\})=- \sum_{m,n} q_{mn} \log q_{mn} $ and  $m,n=0,1,...,d-1$. 
Here, the subscripts  \textquotedblleft \textmd{un}\textquotedblright, 
\textquotedblleft \textmd{B}\textquotedblright, \hspace{0.2mm}  and the 
superscript \textquotedblleft \textmd{P}\textquotedblright \hspace{0.2mm} refer to unitary encoding,  Bell states, and Pauli channels, respectively.
Notice that the super dense coding capacity of a $d$-dimensional Bell state
in the  noiseless case is given by $\log d^2$.
We can then see that in the presence of a one-sided Pauli channel
the super dense coding capacity is reduced by the amount $H(\{q_{mn}\})$
with respect to the noiseless case.
Notice that the same capacity is achieved also for any maximally
entangled state, i.e. for any $\ket{\psi}=U_a\otimes U_b\ket{\psi_{00}}$.

Let us now consider an input Werner
state $\rho_w=\eta\rho_{00}+(1-\eta)\frac{\id}{d^2}  $ with $0\leq \eta \leq 1$ and $\rho_{00}= \ket{\psi_{00}}\bra{\psi_{00}}$. 
For this case it was also shown that  the output von Neumann entropy is 
independent of the unitary encoding $U_{min}$ 
and it is equal to the Shannon entropy $H(\{\frac{1-\eta}{d^2}+\eta q_{mn}\})$ \cite{zahra-paper}.  Therefore, according to  eq. (\ref{dc-covariant}),  the super dense coding capacity  for an input Werner state is given by

\begin{eqnarray}
C_{\mathrm{un,w}}^{\mathrm{P,} \mathrm{one-sided}}=\log d^2-H(\{\frac{1-\eta}{d^2}+\eta q_{mn}\}).
\label{werner-onesided-pauli}
\end{eqnarray}
Here, the subscript \textquotedblleft \textmd{w}\textquotedblright, \hspace{0.2mm} refers to a Werner state. We also notice that, for a Werner state,  the reduced state is  $\rho_b=\frac{\id^{\textmd{}}}{d}$. The above capacity is also achieved by any other state with the form
$U_a\otimes U_b \rho_w U^\dagger_a\otimes U^\dagger_b$.
\subsection{Two-sided channel \label{two-sid}}

In this section we consider the case where both Alice's and Bob's
subsystems undergo covariant noise, described by eq. (\ref{2-Pauli-channels}).
We will first consider the case of the uncorrelated \emph{d}-dimensional depolarising
channel, where the probabilities are given by

\begin{eqnarray}
q_{mn}=
\begin{cases}
  1-p+\frac{p}{d^2},  & m=n=0 \\
  \frac{p}{d^2}, & \mbox{otherwise}
\end{cases}
\label{qmn}
\end{eqnarray}
\vspace{0.5cm}
for the noise parameter $p$, $0\leq p \leq 1$ and $m,n=0,...,d-1$.

For this particular form of Pauli channel and with the probability  $ q_{mn\tilde{m}\tilde{n}} = q_{mn}  q_{\tilde{m}\tilde{n}} $, it can be easily proved
\cite{zahra-paper} that for a state $\rho$ and bilateral
unitary operator $U_a\otimes U_b$, we have
 \begin{eqnarray}
S\left(\Lambda_{ab}^{\mathrm{dep}} \left(\left(U_a\otimes U_b\right)\rho
(U_a^\dagger\otimes U_b^\dagger )\right)\right)=
S(\Lambda_{ab}^{\mathrm{dep}}(\rho)).
\label{depo-unitaryinvariant}
\end{eqnarray}
Therefore the von Neumann entropy of the output
state does not depend on the encoding procedure
that was performed before the action
of the channel. Therefore, according to  eqs. (\ref{dc-covariant}) and (\ref{depo-unitaryinvariant}), the super dense coding capacity
then takes the simple form

\begin{eqnarray}
C^{\mathrm{two-sided}}_{\mathrm{dep}}(\rho)
&=& \log d +S\left(\Lambda_b^{\mathrm{dep}}\left(\rho_b\right)\right)-S\left(\Lambda_{ab}^{\mathrm{dep}}\left(\rho\right)\right).\nonumber\\
\label{DC-2usedepo-any-state}
\end{eqnarray}

Notice that, since eq. (\ref{depo-unitaryinvariant})
holds for any local unitary $U_a\otimes U_b$,
the capacity (\ref{DC-2usedepo-any-state}) depends only on the degree
of entanglement of the input state $\rho$. In other words, all input states
with the same degree of entanglement have the same super dense coding capacity.
The above expression must be compared with the one related to the noiseless
case, given by $C=\log d+S(\rho_b)-S(\rho)$ \cite{ourPRL}.

We can then study how eq. (\ref{DC-2usedepo-any-state}) can be maximised as
a function of the input state for the case of two qubits ($d=2$) and
then derive the optimal value of the dense coding capacity for the qubit
depolarising channel. By following the approach of \cite{zahra-paper},
it turns out that the optimal
capacity for the two-sided qubit depolarising channel is a non differentiable
function of the noise parameter $p$, and
the optimal states are either maximally entangled or separable. In other words,
there is a transition in the entanglement of the optimal input states
at the particular threshold value of the noise parameter
$p_t=0.345$. Notice that a similar
transition behaviour in the entanglement of the optimal input states for
transmission of classical information was found also for the qubit
depolarising channel with correlated noise \cite{mp}.
It is interesting that in the present context the transition behaviour arises
in a memoryless channel and is not related to correlations introduced by
the noise process.

The resulting capacity has also been compared to the classical capacity
for transmission of a single qubit through a depolarising channel, derived in
\cite{king}, as shown in Fig. \ref{fig1}.

\newpage
.

\vspace{2cm}

\setlength{\unitlength}{1cm}

\begin{figure}[h]
\begin{picture}(9,2)
\put(2.8,0.7){\includegraphics[scale=0.70,bb=4cm 2cm 8cm 8cm]{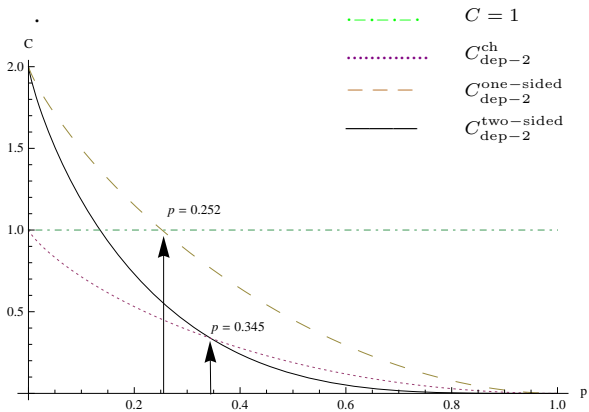}}

\put(4.5,4.5)  {${\color{green}.}$${\color{green}_{-}}${\color{green}.}$\color{green}_{-}$\color{green}.$\color{green}_{-}$$\color{green}.$}

\put(6,4.5){ {\scriptsize $ C=1$}}

\put(4.5,4){\color{violet}..............}
\put(6,4){ {\scriptsize $ C^{\mathrm{ch}}_{\mathrm{dep}-2}$}}

\put(4.5,3.5){$\color{brown}-$    $\color{brown}-$      $\color{brown}-$}
\put(6,3.5){ {\scriptsize  $C^{\mathrm{one-sided}}_{\mathrm{dep-2}}$ }
}
\put(4.5,3){---------}
\put(6,3){ {\scriptsize  $C^{\mathrm{two-sided}}_{\mathrm{dep-2}}$}
}
\end{picture}
\vspace{0.7cm}
\caption{
(Color online) Classical capacity $ C=1$, two-dimensional depolarising
channel capacity  $ C^{\mathrm{ch}}_{\mathrm{dep-2}}$  and super dense coding
capacities with a Bell state in the presence of a {one-sided} and
{two-sided} 2-dimensional depolarising channel,
$C^{\mathrm{one-sided}}_{\mathrm{dep-2}}$ and
$C^{\mathrm{two-sided}}_{\mathrm{dep-2}}$ respectively, as functions of 
the noise parameter $p$.
}
\label{fig1}
\end{figure}

We will now consider the case of a correlated channel, and summarise
in particular
the case of a $d$-dimensional quasi-classical depolarising channel (or simply
quasi-classical channel \cite{memory-quasi-chiara,cerf}), and the case of
a fully correlated Pauli channel, which were extensively reported in Ref.
\cite{zahra-mem}. For the quasi-classical channel, the
probabilities of  the { displacement }operators $V_{mn}$ are equal for $m=0$
and any phase shift labelled by $n$,
and they differ from the rest of the probabilities which are also equal, i.e.
\begin{eqnarray}
q_{mn}=
\begin{cases}
  \frac{1-p}{d},  & m=0 \\
  \frac{p}{d(d-1)}, & \mbox{otherwise}.
\end{cases}
\label{qmn-quasiclassica}
\end{eqnarray}
 The quasi-classical  channel is characterised by a single probability
parameter $ 0 \leq p\leq 1$. With probability $p$,
a {displacement} occurs and with probability $1-p$, no
{displacement} occurs to the quantum signal. Like in the
classical case, $p$ can also be seen as the amount of noise in a
channel.

In the following we will consider as a resource state a Werner state
$\rho_{\textmd w}=  \eta
\ket{\varPhi^+}\bra{\varPhi^+}+(1-\eta)\frac{\id} {4}$ with
$\ket{\varPhi^{+}}=\frac{1}
{\sqrt{2}}\left(\ket{00}+\ket{11}\right)$.  Thus the dimension is $d=2$.
The {correlated} quasi-classical  channel is
in this case
\begin{eqnarray}
\Lambda^Q_{\textmd {ab}} (\xi)= \sum_{m,n} q_{mn} \sigma_m\otimes\sigma_n (\xi)\sigma_m\otimes\sigma_n,
\label{Q-ab}
\end{eqnarray}
where $q_{mn}=  (1-\mu) q_m q_n+ \mu q_n\delta_{mn} $ with
$q_0=q_3=\frac{1-p}{2}$ and $q_1=q_2=\frac{p}{2}$, while $\sigma_0=I$, while
$\sigma_1$, $\sigma_2$ and $\sigma_3$ denote the Pauli operators along
axes $x,y,z$,  respectively. In \cite{zahra-mem} , for a  Werner state in a {correlated} quasi-classical channel, we proved that  $U_{\textmd{min}}=I$. Therefore, for this case, according to  eq. (\ref{dc-covariant}),  the super dense coding capacity is given by
\begin{eqnarray}
C_{\textmd{un}}^{\textmd{Q,w}}=2-S\left(\Lambda^Q_{\textmd {ab}}\left(\rho_{\textmd w} \right)\right).
\label{werne-quasiclassical}
\end{eqnarray}

For $\eta=1$, the Werner state $\rho_{\textmd w}$ reduces to a Bell state
$\ket{\varPhi^+}$. Therefore, the super dense coding capacity, according to
(\ref{werne-quasiclassical}), for a Bell state and in the presence of a
{correlated} quasi-classical channel, is given by
\begin{eqnarray}
C_{\textmd{un}}^{\textmd {Q,B}}=2-S\left(\Lambda^Q_{\textmd {ab}}
\left( \ket{\varPhi^+}\bra{\varPhi^+}\right)\right).
\label{C-un-Bell}
 \end{eqnarray}

\begin{figure}[h]
\begin{center}
\begin{tabular}{c  }
\includegraphics[width=8.5 cm]{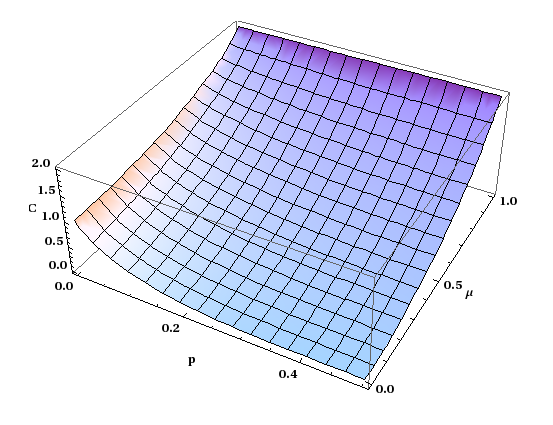}  \\
\end{tabular}
\end{center}
\caption{ The super dense coding capacity for a
correlated quasi-classical channel  and a Bell state ($\eta=1$ ), as a
function of the noise parameter $p$ and the correlation degree $\mu$.}
\label{fig2}
\end{figure}

\begin{figure}[h]
\begin{tabular}{c  }
\includegraphics[width=8.5 cm]{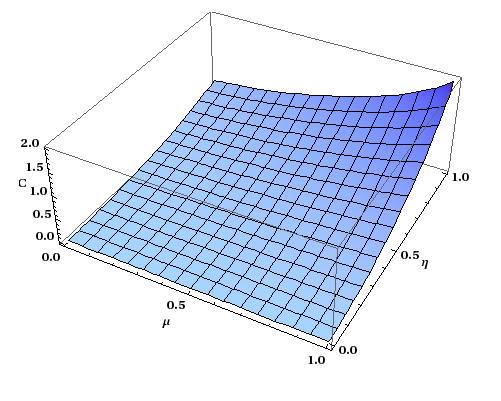}  \\
\end{tabular}
\caption{The super dense coding capacity for a correlated quasi-classical
channel and a Werner state, as a function of the correlation degree $\mu$ and
the  parameter $\eta$. The noise parameter is $p=0.05$.}
\label{fig3}
\end{figure}

In Fig. \ref{fig2} and Fig. \ref{fig3}, we report the super dense coding capacity for the
{correlated}  quasi-classical channel as a function of the parameters $\mu$,
$\eta$ and $p$  (eqs. \ref{Q-ab}, \ref{werne-quasiclassical}). In Fig. \ref{fig2}, we consider a
Bell state, i.e. $\eta=1$,  as a function of the noise parameter $p$ and the
correlation degree $\mu$ (eqs. \ref{Q-ab}, \ref{C-un-Bell}).
In Fig. \ref{fig3}, we fix the noise parameter to $p=0.05$ and we vary $\mu$ and the
parameter $\eta$ characterising the Werner state.

\begin{table*}
  \centering
\begin{tabular}{ |  c ||    c  | c  |   c  | c|  c| c  |}\hline
\diaghead{\theadfont Diag ColumnmnHead II}%
{ \\Resource state}{  \\ Channel } & \thead{  One-sided  \\  $d$-dimensional\\Pauli channel  }    & 
\thead{  Two-sided   \\ $d$-dimensional  \\ depolarising channel \\ with $\mu=0$ } & \thead{  Two-sided   \\ $d$-dimensional  \\ correlated depolarising channel\\with the correlation degree  $\mu$  }& \thead{ Two-sided \\  $2$-dimensional  \\ correlated quasi-classical channel \\with the correlation degree  $\mu$  } &\thead{ Two-sided \\  $2$-dimensional  \\ fully-correlated \\ Pauli channel ($\mu=1$)  }\\
\hline\hline

Bell state& eq. (\ref{DCcapacityd-qubitcha})  &  eq. (\ref{DC-2usedepo-any-state})  & {\color{red}open}& eq. (\ref{C-un-Bell}) &   $C=2$ \\  \hline

Werner state&eq. (\ref {werner-onesided-pauli})  &  eq. (\ref{DC-2usedepo-any-state}) & {\color{red}open}& eq. (\ref{werne-quasiclassical}) & eq. (\ref{c-fully-werner})  \\  \hline

Bell diagonal state& {\color{red}open}  &  eq. (\ref{DC-2usedepo-any-state}) & {\color{red}open}& {\color{red}open} & eq. (\ref{Bell-diagonal-c})  \\  \hline

Arbitrary state $\rho$ &{\color{red}open}  &  eq. (\ref{DC-2usedepo-any-state}) &{\color{red}open} &{\color{red}open}  &  {\color{red}open} \\  \hline

\end{tabular}
\caption{A summary of the solved examples of  bipartite resource states and channels for super dense coding capacity with unitary encoding. Here, some of the unsolved (open) examples are also mentioned.}
\label{table1}
\end{table*}

As mentioned above, we now consider the case of a {fully correlated} Pauli
channel, which is a special form
of a {correlated} Pauli channel where
$\mu=1$.  For $d=2$ it is given by
\begin{eqnarray}
\Lambda^{\textmd f}_{\textmd {ab}}(\xi)
&=&\sum_{m}q_m(\sigma_m\otimes \sigma_m)(\xi)(\sigma_m\otimes \sigma_m),
\label{fully-pauli}
\end{eqnarray}
where $\sum_{m}q_m =1$ and $\sigma_m$ are either the identity or the Pauli
operators.

{We  consider Bell diagonal states $\rho_{Bd} $ as resource states. A Bell diagonal state is  a convex combination of the four Bell states. That is $ \rho_{Bd} = \sum^3_{n=0}  p_n\rho_n $, where $\rho_n$ is a Bell state, $p_n  \geq  0$,
and  $\sum^3_{n=0} p_n = 1$.
The subscript    \textquotedblleft \textmd{Bd}\textquotedblright \hspace{0.2mm} stands for a Bell
diagonal state. For a fully correlated Pauli  channel (\ref{fully-pauli}) we 
now determine the unitary operation $U_{\textmd{min}}$. To do so,  since applying a unital CPTP map cannot decrease  the von Neumann entropy \cite{ent-unit},  
and since the von Neumann entropy is invariant under applying the unitary operator,  we find the following  lower bound  on   $S\left(\Lambda^{\textmd f}_{\textmd {ab}}\left((U\otimes\id)\rho_{\textmd Bd}
(U^\dagger\otimes\id)\right)\right)$

\begin{eqnarray}
 S\left(  \rho_{\textmd Bd} \right)  \leq S\left(\Lambda^{\textmd f}_{\textmd {ab}}\left((U\otimes\id)\rho_{\textmd Bd}
(U^\dagger\otimes\id)\right)\right) ,
\label{bell-diag-lowerbound}
\end{eqnarray}
where $U$ is an arbitrary
unitary operator, and  $\Lambda_{\textmd {ab}}^\textmd{f}$ is the
unital map for the fully correlated Pauli channel. By using the invariance of a Bell state under the action of a
{fully correlated} Pauli channel, i.e., $\Lambda_{\textmd {ab}}^\textmd{f}
(\rho_n)=\rho_n$, it follows that the lower bound
(\ref{bell-diag-lowerbound}) is reachable by
 the identity operator. Then
$U_{\textmd{min}}=\id$ and the super dense coding capacity, according to  eq. (\ref{dc-covariant}), is given by
\begin{eqnarray}
C_{\textmd{un}}^{\textmd{f, Bd}}=2-S\left(\rho_\textmd{Bd} \right).
\label{Bell-diagonal-c}
\end{eqnarray}  
The capacity (\ref{Bell-diagonal-c}) shows that for a fully correlated  Pauli channel and a Bell diagonal state,  this class of channels
behaves like a noiseless one.

The Werner state $\rho_{\textmd w}=  \eta
\ket{\varPhi^+}\bra{\varPhi^+}+(1-\eta)\frac{\id} {4}$ is a special case of  a Bell diagonal state with $p_0= \frac{1+3\eta} {4}$ and $p_1=p_2=p_3= \frac{1-\eta} {4}.$ Therefore, according to  eq.  (\ref{Bell-diagonal-c}), the super dense coding capacity for a Werner state in the presence of a fully correlated Pauli channel is given by 

\begin{eqnarray}
C_{\textmd{un}}^{\textmd{f,w}}=
2-S\left(\rho_\textmd{w} \right).
\label{c-fully-werner}
\end{eqnarray}}

The Werner state $\rho_{\textmd w}$ reduces to a Bell state
$\rho^+= \ket{\varPhi^+} \bra{\varPhi^+}$
for $\eta=1$. Since the Bell state is a pure state, its  von Neumann entropy
   $S\left(\rho^+ \right)$
is zero.  Therefore, using (\ref{c-fully-werner}), the super dense coding
capacity for a shared Bell state and a fully correlated Pauli channel is {two} bits. It is the
 maximum information transfer for $d=2$. This shows that no information at
all is lost to the
environment and this class of channels behaves like a noiseless one.
This behaviour corresponds to the results of \cite{mp,memory-quasi-chiara}.

{In  Table \ref{table1},  we summarise all the  bipartite solved examples given in this section. We also mention some of the unsolved cases as \textquotedblleft open\textquotedblright . }


\section{Non-unitary encoding versus unitary encoding \label{s:nu}}

{In section \ref{s:pauli},  we provided some examples of resource
states and channels, for which we derived the capacity with unitary encoding. For all those examples the optimal 
$U_{\textmd{min}}$ is the \textquotedblleft identity operator\textquotedblright.  The problem of finding the non-unitary pre-processing $\Gamma_{\textmd{min}}$ is obviously  more difficult  than  finding 
$U_{\textmd{min}}$, as the optimisation runs over all CPTP maps.} The application of an appropriate pre-processing
$\Gamma_{\textmd{min}}$ on the initial state $\rho$ before the unitary
encoding $\{V_i\}$  may increase the super dense coding capacity
with respect to only using unitary encoding for the case of a covariant
channel. However, for some examples no  better encoding than unitary encoding
is possible. For instance, since \textit{two} bits  is the highest super
dense coding capacity for $d=2$, the results derived in the previous section
for a {fully correlated} Pauli channel and the  Bell state provide  an example
where no pre-processing can improve the capacity.
However, there exist
examples for which non-unitary pre-processing is useful to increase the super
dense coding capacity. One of these examples is the case of a two-sided
depolarising channel for qubits, discussed in the previous section.
Actually, consider this channel with a Bell state with  noise parameter in
the range  $0.345\leq p \leq 1$.
In this case, before performing the encoding, Alice applies a
measurement as a ``pre-processing'' in order to project the Bell state onto
$\ket{00}$ or $\ket{11}$  and then applies the unitary encoding.
As we mentioned before, the super dense coding capacity for product states
is equal to the depolarising channel capacity, shown in Fig. \ref{fig1}.
In this case we then reach the 2-dimensional depolarising channel capacity,
which is higher than the super dense coding capacity without pre-processing,
as shown in Fig. \ref{fig1}.

\begin{figure}[h]
\begin{center}
\begin{tabular}{c  }
\includegraphics[width=8 cm]{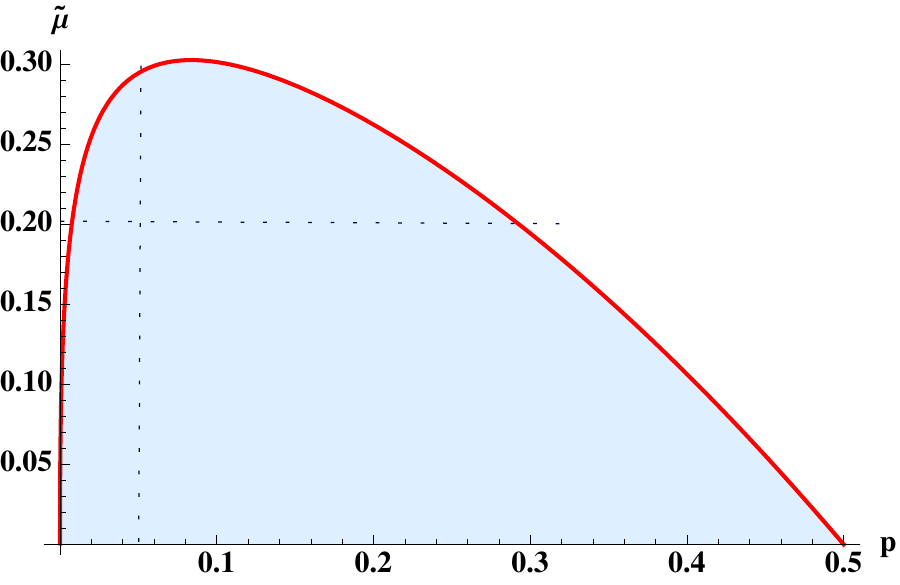}  \\
\end{tabular}
\label{mu-p}
\end{center}
\caption{ (Color online) The \textquotedblleft red\textquotedblright \hspace{0.7mm} curve is the correlation degree $\tilde{\mu}(p)$ as a function of
the noise parameter $p$. The super dense coding capacity $C^{\textmd{Q,B}}_{\textmd {un}}$ and the transferred information $C^{\textmd{Q,B}}_{\Gamma}$ coincide for $\mu=\tilde{\mu}(p)$ (see main text). For $\mu < \tilde{\mu}(p) $, the
\textquotedblleft blue\textquotedblright \hspace{0.7mm} area, the non-unitary pre-processing $\Gamma$ increases the super dense coding
capacity of a quasi-classical channel and a Bell state, in comparison to just
unitary encoding. The horizontal and vertical lines correspond to the parameters   ${\mu}=0.2$ and  $p=0.05$,  chosen in Fig. \ref{fig5} and Fig. \ref{fig6}, respectively.}
\label{fig4}
\end{figure}

\begin{figure}[h]
\begin{center}
\begin{tabular}{c  }
\includegraphics[width=7.5cm]{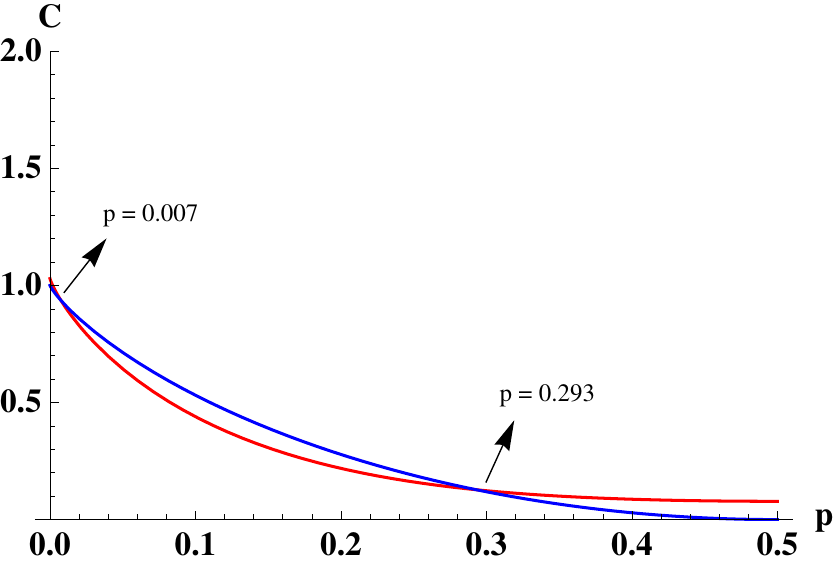}  \\
\end{tabular}
\label{mu-p}
\end{center}
\caption{(Color online) Comparison between the super dense coding capacity (\ref{C-un-Bell}), and the transferred information (\ref{non-unitary-qusi-Bell}) as functions of the noise parameter $p$, for an example with the correlation degree ${\mu}=0.2$. The \textquotedblleft red\textquotedblright \hspace{0.7mm} curve corresponds to the capacity $C^{\textmd{Q,B}}_{\textmd {un}}$ given by eq. (\ref{C-un-Bell}), while the  \textquotedblleft blue\textquotedblright \hspace{0.7mm} curve represents  the transferred information $C^{\textmd{Q,B}}_{\Gamma}$ given by eq. (\ref{non-unitary-qusi-Bell}). As we can see from the above graph, for ${\mu}=0.2$, in the range of the noise parameter $ 0.007 < p < 0.293 $, we reach a higher capacity by applying the non-unitary pre-processing $\Gamma$, the \textquotedblleft blue\textquotedblright \hspace{0.7mm} curve.}
\label{fig5}
\end{figure}

\begin{figure}[h]
\begin{center}
\begin{tabular}{c  }
\includegraphics[width=6cm]{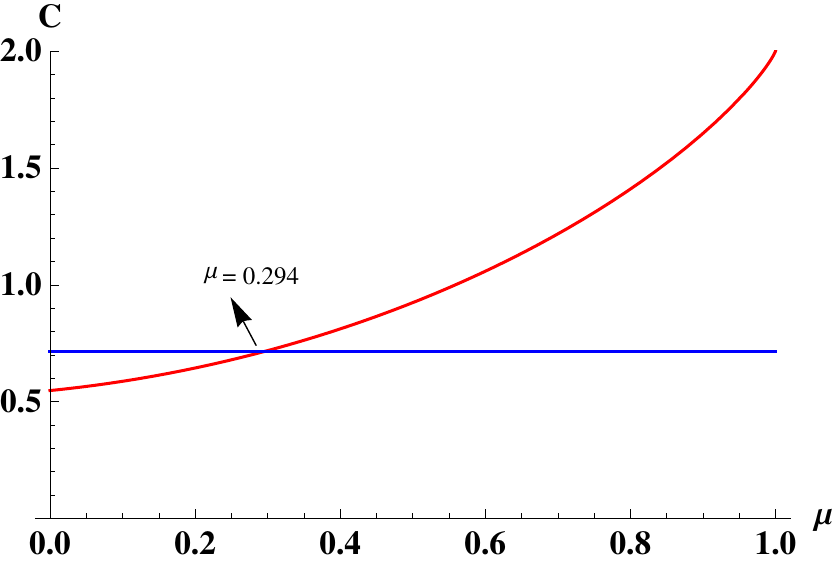}  \\
\end{tabular}
\label{mu-p}
\end{center}
\caption{(Color online)  Another comparison between the super dense coding capacity (\ref{C-un-Bell}), and the transferred information (\ref{non-unitary-qusi-Bell}) as functions of the correlation degree $\mu$, for an example with the noise parameter $p=0.05$. The \textquotedblleft red\textquotedblright \hspace{0.7mm} curve corresponds to the capacity $C^{\textmd{Q,B}}_{\textmd {un}}$ given by eq. (\ref{C-un-Bell}). The  \textquotedblleft blue\textquotedblright \hspace{0.7mm} line represents the transferred information $C^{\textmd{Q,B}}_{\Gamma}$ given by eq. (\ref{non-unitary-qusi-Bell}). As we can see from the above graph, for $p=0.05$, and for $ \mu < 0.294$, the non-unitary pre-processing $\Gamma$ is useful to enhance the capacity, compared to only unitary encoding.}
\label{fig6}
\end{figure}

As another example, consider a two-dimensional Bell state in the presence of a
correlated quasi-classical channel. In this case a non-unitary
pre-processing $\Gamma$, which is not necessarily $\Gamma_{min}$,
can improve the super dense coding capacity. To show this claim, consider
the completely positive trace preserving pre-processing $\Gamma$, with the
Kraus operators $E_1=\ket{0}\bra{1}$ and $E_2=\ket{0}\bra{0}$.
Alice applies $\Gamma$ on her side of the Bell state
$\rho^+=\ket{\varPhi^+}\bra{\varPhi^+}$ and transforms the Bell state to
$\Gamma (\rho^+) =\ket{0}\bra{0} \otimes \frac{\id}{2}$.
Therefore, according to eq. (\ref{capacity-non-un}), for a correlated
quasi-classical channel as a special case of a covariant channel, a Bell state, and a pre-processing $\Gamma$, the
amount of information that is transmitted by this process is given by
\begin{eqnarray}
C^\textmd{{Q,B}}_{\Gamma}=1+p \log p + (1-p) \log (1-p),
\label{non-unitary-qusi-Bell}
\end{eqnarray}
where $p$ is the noise parameter for a quasi-classical channel
(\ref{qmn-quasiclassica}). Since $\Gamma$ is not necessarily the optimal
pre-processing, $ C^\textmd{{Q,B}}_{\Gamma}$ is not also necessarily the
capacity.
We name (\ref{non-unitary-qusi-Bell}) the \emph{transferred information}.
We now compare the {transferred information} (\ref{non-unitary-qusi-Bell})
with the capacity (\ref{C-un-Bell}) which is achieved by applying only unitary
encoding.
In the range of $0.3 \leqslant \mu \leqslant 1$
we find that the capacity $C^{Q,B}_{\textmd {un}}$ is always higher than
the  {transferred information} $C^{Q,B}_{\Gamma}$,
i.e. $C^{Q,B}_{\Gamma}\textless C^{Q,B}_{\textmd {un}}$.
Therefore, in this range, the chosen pre-processing $\Gamma$ does not improve
the capacity.
In the range of $0\leqslant \mu \textless 0.3$, the capacity with unitary
encoding (\ref{C-un-Bell}) and the {transferred information} with the
pre-processing $\Gamma$ (\ref{non-unitary-qusi-Bell}) coincide for
$\mu=\tilde{\mu}(p)$, the red curve in Fig. \ref{fig4}.
Note that $\tilde{\mu}(p)$ corresponds to the
$\textmd {Root}[C^{Q,B}_{\textmd {un}}-C^{Q,B}_{\Gamma}]$.
The function $\tilde{\mu}(p)$ is invariant under the simultaneous exchange
$p \leftrightarrow 1-p$ since both functions $C^{Q,B}_{\textmd {un}}$ and
$C^{Q,B}_{\Gamma}$ are symmetric under the exchange $p \leftrightarrow 1-p$.
Our results show that for $\mu\textless\tilde{\mu}(p)$, the
\textquotedblleft blue\textquotedblright \hspace{0.7mm} area in Fig. \ref{fig4},
the {transferred information} (\ref{non-unitary-qusi-Bell}) leads to a
higher value, in comparison to the capacity given by eq.(\ref{C-un-Bell}),
i.e. $C^{Q,B}_{\Gamma}\textgreater C^{Q,B}_{\textmd {un}}$.
In Fig. \ref{fig5}, and Fig. \ref {fig6}, we plot the super dense coding capacity
corresponding to unitary encoding and the  {transferred information}
corresponding to the pre-processing $\Gamma$, eqs. (\ref{C-un-Bell}) and
(\ref{non-unitary-qusi-Bell}).
In Fig. \ref{fig5}, the correlation degree is ${\mu}=0.2$, while we vary the noise
parameter $p$. In Fig. \ref{fig6}, the noise parameter is $p=0.05$ and $\mu$ is varied.

{{Considering unitary encoding, in Section (\ref{two-sid}),  eq. (\ref{c-fully-werner}),  for a Werner state  in the presence of a fully correlated Pauli channel, the super dense coding capacity was derived.  We notice that the capacity is just characterised by the Werner state and  it is independent 
of the noise parameters that characterise the fully correlated Pauli channel. Here, as the last example,  we show that a non-unitary pre-processing $\Gamma$ with the Kraus operators $E_1=\ket{0}\bra{1}$ and $E_2=\ket{0}\bra{0}$, introduced  also in the previous example,  can once more enhance the capacity in comparison to eq. (\ref{c-fully-werner}). To show this claim,  we  apply $\Gamma$ on Alice's side of the  Werner state $\rho_{\textmd w}=  \eta
\ket{\varPhi^+}\bra{\varPhi^+}+(1-\eta)\frac{\id} {4}$ and so we  map it to  the state $\Gamma (\rho_{\textmd w}) =\ket{0}\bra{0} \otimes \frac{\id}{2}$. Therefore,  according to eq. (\ref{capacity-non-un}),  for a fully correlated Pauli channel,  a Werner state, and the pre-processing $\Gamma$, the
amount of information that is transmitted by this process is given by

\begin{eqnarray}
C_{\Gamma}^{\textmd{f,w}}&=&2-S\left(\Lambda^f_{\textmd {ab}}\left(   \Gamma(\rho_{\textmd w} )\right)\right).\nonumber\\
&=&2- S\left(\Lambda^f_{\textmd {ab}}\left(  \ket{0}\bra{0}\otimes \frac{\id}{2}\right)\right).\nonumber\\
&=&  2- S\Big((q_0+q_3)\ket{0}\bra{0}+(q_1+q_2)\ket{1}\bra{1}\Big)-S( \frac{\id}{2})  \nonumber\\
&=&1+q \log q + (1-q) \log (1-q),
\label{wer-f-gamma}
\end{eqnarray}
where in the last line we have used the new notation $q:=q_0+q_3$ and so $ 1-q :=q_1+q_2$.  Since $\Gamma$ is not necessarily the optimal
pre-processing, $C^\textmd{{f,w}}_{\Gamma}$ is not also necessarily the
capacity. Similar to the previous example, we name  (\ref{wer-f-gamma}) the \emph {transferred information}.   We notice that  the transferred information  $C_{\Gamma}^{\textmd{f,w}}$ is characterised only by  the channel 
parameter $q$ and it  is invariant under the exchange
$q \leftrightarrow 1-q$.  We now provide a comparison between  the capacity (\ref{c-fully-werner}) and  the transferred information (\ref{wer-f-gamma}). In Fig. \ref{c-eta-q}, these two functions   versus $q$ and $\eta$ are depicted in a three dimensional plot. The \textquotedblleft green\textquotedblright \hspace{0.7mm} curve stands for the capacity $C^{\textmd{f,w}}_{\textmd {un}}$, while  the \textquotedblleft blue\textquotedblright \hspace{0.7mm} curve represents the transferred information $C^\textmd{{f,w}}_{\Gamma}$. We can prove that for $0.747<\eta\leq 1 $,  the capacity $ C^{\textmd{f,w}}_{\textmd {un}}$, the green curve, is always higher than the transferred information $C^\textmd{{f,w}}_{\Gamma}$, the blue curve, for all values of $q$. Accordingly,  the pre-processing $\Gamma$ cannot increase the capacity in this range of $\eta$. However, in the range $0\leq\eta\ < 0.747$,  depending on values of $q$ and $\eta$, the pre-processing $\Gamma$  can  increase the capacity.  Our results show that the two functions $C_{\Gamma}^{\textmd{f,w}}$ and $C^{\textmd{f,w}}_{\textmd {un}}$ coincide for $\eta=\tilde{\eta}(q)$, reported as 
the \textquotedblleft red\textquotedblright \hspace{0.7mm} curve  in Fig. \ref{etatilde-q}. We can also show that for $\eta < \tilde{\eta}(q) $, the \textquotedblleft blue\textquotedblright \hspace{0.7mm} area in Fig. \ref{etatilde-q},  the transferred information  (\ref{wer-f-gamma}) is higher than the capacity  with unitary encoding (\ref{c-fully-werner}), i.e. 
 $C^{\textmd{f,w}}_{\textmd {un}} < C_{\Gamma}^{\textmd{f,w}}$.  Therefore, for the blue area in  Fig. \ref{etatilde-q},  applying  the non-unitary pre-processing $\Gamma$ leads to a higher information transfer  for a Werner state and a fully correlated Pauli channel in comparison to just unitary encoding.

\begin{figure}[h]
\begin{center}
\begin{tabular}{c  }
\includegraphics[width=8cm]{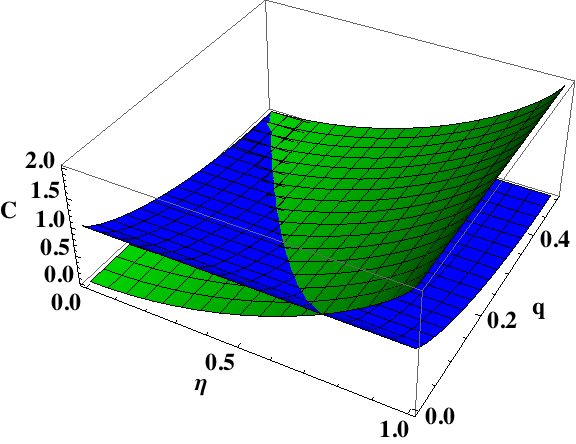}  \\
\end{tabular}
\end{center}
\caption{(Color online) Comparison between the super dense coding
capacity  with unitary encoding $C^{\textmd{f,w}}_{\textmd {un}}$, green curve,  and the transferred information $C^{\textmd{f,w}}_{\Gamma}$, blue curve,  as functions
of  $\eta$ and $q$,  for a Werner state and a fully correlated Pauli channel. } 
\label{c-eta-q}
\end{figure}}

\begin{figure}[h]
\begin{center}
\begin{tabular}{c  }
\includegraphics[width=8cm]{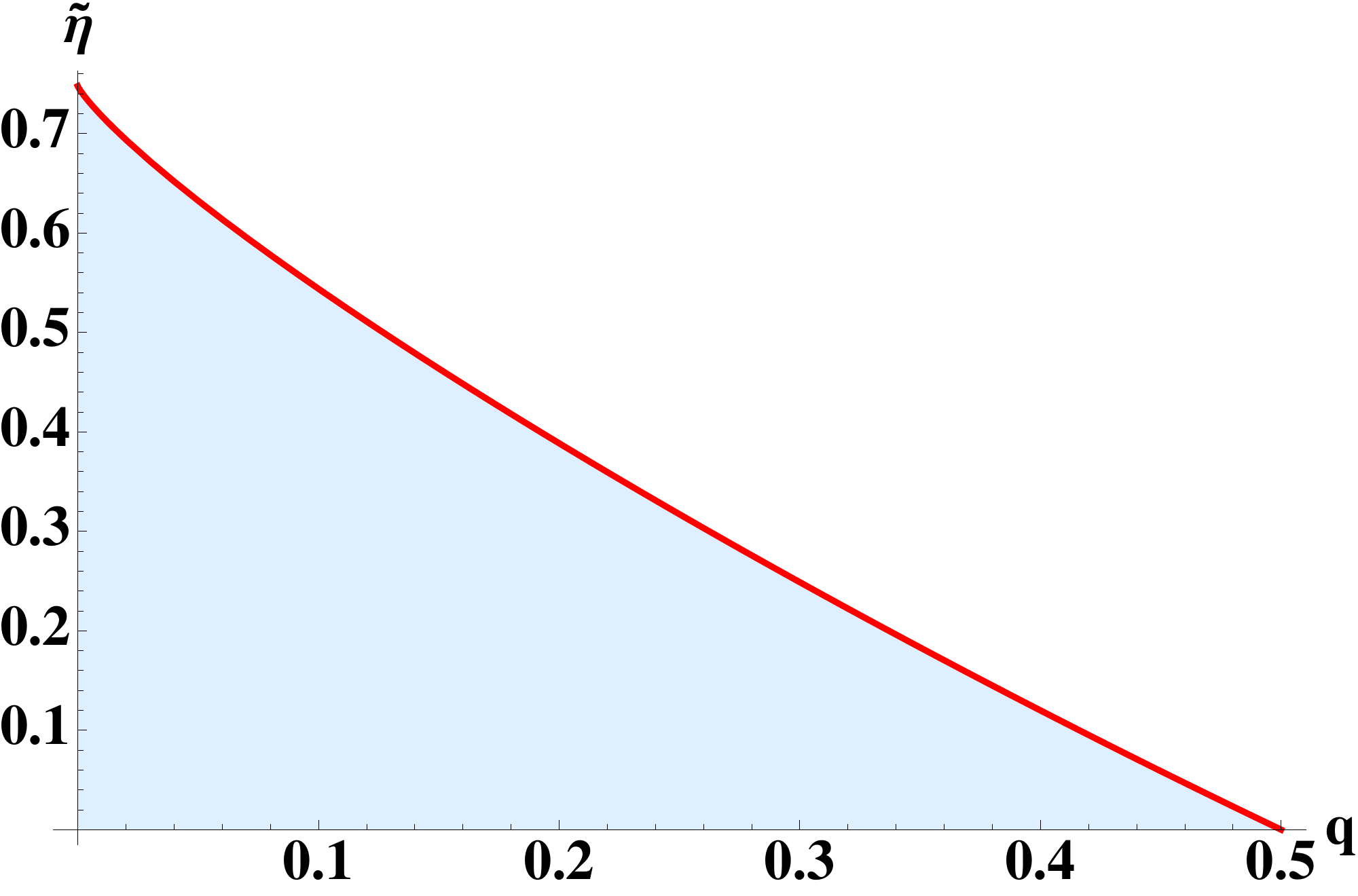}  \\
\end{tabular}
\end{center}
\caption{ (Color online) The \textquotedblleft red\textquotedblright \hspace{0.7mm} curve shows $\tilde{\eta}(q)$  as a function of
the channel parameter $q:=q_0+q_3$. The super dense coding capacity $C^{\textmd{f,w}}_{\textmd {un}}$ and the transferred information $C^{\textmd{f,w}}_{\Gamma}$ coincide for $\eta=\tilde{\eta}(q)$ (see main text). For $\eta < \tilde{\eta}(q) $, the
\textquotedblleft blue\textquotedblright \hspace{0.7mm} area, the non-unitary pre-processing $\Gamma$ increases the super dense coding
capacity of a fully correlated Pauli  channel and a Werner state, in comparison to just
unitary encoding. }
\label{etatilde-q}
\end{figure}

\section{Distributed super dense coding \label{s:multi}}

The multipartite super dense coding scheme  works as follows:
a given quantum state $\rho^{\textmd {a}_1...\textmd {a}_k \textmd {b}}$
is distributed between $k$ Alices and a single Bob (in our scenario,
Bob's subsystem experiences  noise in this stage). Then,  Alices  perform
with the  probability $p_{\{i\}}$ a unitary operation
$W_{\{i\}}^{\textmd {a}_1...\textmd {a}_k}$ on their side of the state
$\rho^{\textmd {a}_1...\textmd {a}_k\textmd {b}}$,  thus  encoding  classical
information through the state  $\rho_ {\{i\}}=(W_ {\{i\}}^{\textmd
{a}_1...\textmd {a}_k} \otimes \id^b) \hspace{2mm} \rho^{\textmd
{a}_1...\textmd {a}_k\textmd {b}} \hspace{2mm}
(W_{\{i\}} ^{ \textmd {a}_1,...,\textmd {a}_k\dagger} \otimes \id^\textmd
{b})$, where $\id^\textmd {b}$  is the identity operator on the  Bob's
Hilbert space and ${\{i\}}$, is a set of indices for Alices. Subsequently,
the Alices send their subsystems of the encoded state  through the noisy
channel to Bob. We consider
$\Lambda_{\textmd {a}_1...\textmd {a}_k\textmd {b}}:\rho_  {\{i\}}\rightarrow \Lambda_{\textmd {a}_1...\textmd {a}_k\textmd {b}}(\rho_ {\{i\}})$ to be the CPTP map (quantum channel) that
globally acts on the multipartite  encoded  state $\rho_  {\{i\}}$.
By this process, Bob receives the ensemble $\{\Lambda_{\textmd {a}_1...
\textmd {a}_k\textmd {b}}(\rho_ {\{i\}}) , p_{\{i\}}\}$.
By performing suitable measurements,  Bob can  extract the  accessible
information about  this ensemble which is  given by the Holevo quantity
\cite{ Holevo-chi-quantity}

\begin{eqnarray}
\chi_{\textmd {un}} \left(\{\rho_{\{i\}},p_{\{i\}}\}\right)&=&S\Big (
\sum_{\{i\}} p_{\{i\}}
 \Lambda_{\textmd {a}_1...\textmd {a}_k\textmd {b}}\left(\rho_{\{i\}}  \right)\Big)\nonumber\\
&-&\sum_{\{i\}} p_{\{i\}} S\left(\Lambda_{\textmd {a}_1...\textmd {a}_k\textmd {b}} \left(
\rho_{\{i\}}\right)\right).
 \label{Holevo-multi}
\end{eqnarray}
As before, the subscript
\textquotedblleft \textmd{un}\textquotedblright \hspace{0.2mm} refers to
unitary encoding. By a generalisation of the bipartite case considered
in the previous sections, the super dense coding capacity
$C_{\textmd {un}}$
for a given resource state
$\rho^{\textmd {a}_1...\textmd {a}_k\textmd {b}}$ and the noisy channel
$\Lambda_{\textmd {a}_1...\textmd {a}_k\textmd {b}}$ is then defined as

\begin{eqnarray}
&&C_{\textmd{un}}=\max_{\{W_{\{i\}}^{\textmd {a}_1...\textmd {a}_k} , p_{\{i\}}\}}\chi_{\textmd{un}} \left(\{{\rho_{\{i\}}},p_{\{i\}}\}\right).
\label{c-multi-optimization}
\end{eqnarray}

In the following we will consider local covariant channels, denoted by
$\Lambda_{\textmd{a}_1...\textmd{a}_k\textmd{b}}^\textmd {c }$, namely

\begin{eqnarray}
\Lambda^\textmd{c}_{\textmd{a}_1...\textmd{a}_k\textmd{b}}(  \tilde{V}_{\{i\}} \rho   \tilde{V}_{\{i\}}^\dagger) =
  \tilde{V}_{\{i\}} \Lambda^\textmd{c}_{\textmd{a}_1...\textmd{a}_k\textmd{b}}(\rho)   \tilde{V}_{\{i\}}^\dagger,
\label{covariance}
\end{eqnarray}
where we have local unitary operators of the form
   \begin{eqnarray}
  \tilde{V}_{\{i\}}=
V_{i_1}^{\textmd{a}_1}\otimes...\otimes V_{i_k}^{\textmd{a}_k}.
  \label{Vtilde}
  \end{eqnarray}

We can consider the case where the $k$ Alices are far
apart and  they are restricted to local unitary operations and
the case  where the Alices are allowed to  perform
entangled unitary encoding. In the first scenario the $j$th  Alice applies
a local unitary operator $W_{i_j}^{\textmd {a}_j}$
with probability $p_{i_j}$ on her subsystem of the shared state
$\rho^{\textmd {a}_1...\textmd {a}_k\textmd {b}}$.
Then the super dense coding capacity  $C_{\textmd{un}}^{\textmd{lo}}$,
by generalising the procedure described above for the bipartite case,
is given by \cite{zahra-multi}
\begin{eqnarray}
C_{\textmd{un}}^{\textmd{lo}}&=&\log D_\textmd{A}+ S\left(\Lambda_{\textmd{b}}^\textmd{} \left(
\rho_{\textmd{b}}\right)\right)\nonumber\\
&-& S\bigg(\Lambda_{\textmd{a}_1...\textmd{a}_k\textmd{b}}^\textmd{c} \Big(\big(U_{\textmd{min}}^{\textmd{lo}}
 \otimes \id^\textmd{b} \big)  \rho^{\textmd{a}_1...\textmd{a}_k\textmd{b}} \hspace{1mm} \big(   U_{\textmd{min}}^{\textmd{lo}\dagger} \otimes\id^\textmd{b} \big)\Big)\bigg), \nonumber\\
\label{local-dc-covariant}
\end{eqnarray}
where  $D_\textmd{A}=d_{\textmd{a}_1} d_{\textmd{a}_2}...d_{\textmd{a}_k}$
is the dimension of the Hilbert space of  the $k$ Alices,
$\mathrm{\tr}_{\textmd{a}_1...\textmd{a}_k}  \Lambda_{\textmd{a}_1...
\textmd{a}_k\textmd{b}}^\textmd{c} \left(\rho^ {\textmd{a}_1...
\textmd{a}_k\textmd{b}} \right)=\Lambda_{\mathrm{b}}^\textmd{}
\left(\rho_{\mathrm{b}}\right)$ and
$U_{\textmd{min}}^{\textmd{lo}}:=U_\textmd{min}^{\textmd{a}_1}  \otimes... \otimes U_\textmd{min}^{\textmd{a}_k}$
is the unitary operator
that minimises the von Neumann entropy after application of this
unitary operator and  the channel  $\Lambda_{\textmd {a}_1...\textmd {a}_k\textmd {b}}^\textmd {c}$  to the initial
state $\rho^{\textmd {a}_1...\textmd {a}_k\textmd {b}} $.

The second case, where the Alices are allowed  to apply entangled
unitary operators, can be treated in an analogous way,
with the only difference that we now have a global unitary  operator
$U_\textmd{min}^{\textmd{g}}$ which  minimises  the output von Neumann
entropy. We  can then show that the optimal encoding is given by the
ensemble $\{\tilde{U}_{\{i\}}=   \tilde{V}_{\{i\}}
U_\textmd{min}^{\textmd{g}},\tilde{p}_{\{i\}}=\frac{1}{D^2_\textmd{A}} \}$,
and the super dense coding capacity $C_{\textmd{un}}^\textmd{g}$ for this
situation is given by \cite{zahra-multi}
\begin{eqnarray}
&&C_{\textmd{un}}^\textmd{g}=\log D_\textmd{A}+ S\left(\Lambda_{\textmd{b}}^\textmd{} \left(
\rho_{\textmd{b}}\right)\right)  \nonumber\\
&&-S\bigg(\Lambda_{\textmd{a}_1...\textmd{a}_k\textmd{b}}^\textmd{c} \Big( \big(U_\textmd{min}^{\textmd{g}}
\otimes\id^\textmd{b}\big)\rho^{\textmd{a}_1...\textmd{a}_k\textmd{b}}\big(U_\textmd{min}^{ \textmd{g}\dagger}  \otimes\id^\textmd{b} \big)\Big)\bigg).\nonumber\\
\label{global-dc-covariant}
\end{eqnarray}
The difference between the capacities $(\ref{local-dc-covariant})$ and $(\ref{global-dc-covariant})$ is the occurrence of the local and global unitary transformation $U_{\textmd{min}}^{\textmd{lo}}$ and $U_\textmd{min}^{ \textmd{g}}$, respectively.

We will now consider the case of $k+1$ parties in the presence of a Pauli
channel

\begin{eqnarray}
&&\Lambda_{\textmd {a}_1...\textmd {a}_k\textmd {b}}^\textmd {P}\big(\xi\big)
=\sum_{\{m_in_i\}}q_{\{m_in_i\}}
 \Big(V_{m_1n_1}^{\textmd {a}_1} \otimes... \otimes V_{m_kn_k}^{\textmd {a}_k}   \otimes \nonumber\\
&&   V_{m_{k+1} n_{k+1}}^{\textmd {b}} \Big)\xi\left({V_{m_1n_1}^{\textmd {a}_1\dagger}}
 \otimes...\otimes {V_{m_kn_k}^{\textmd {a}_k\dagger}}\otimes V_{  m_{k+1} n_{k+1} }^{\textmd {b}\dagger} \right), \nonumber\\
 \label{k+1-channel}
\end{eqnarray}
where the probabilities $q_{\{m_in_i\}}$ add to one. Here, the
notations $\{m_in_i\}_{i=1}^k$  stand for $k$ Alices and ${m_{k+1} n_{k+1}}$
stands for Bob.

As for the bipartite case, this general model of Pauli channels includes both
the case of memoryless channels, where the Pauli noise acts independently
on each of the $k+1$ parties
and the probabilities $q_{\{m_in_i\}}$ are products of the single party
probabilities $q_{mn}$, or more generally the case where the action of noise
is not independent on consecutive uses but is correlated.
For example, for $k+1$ uses  of a Pauli channel we can define a
{correlated} Pauli channel in the multipartite scenario as follows
\begin{eqnarray}
&&q_{\{m_in_i\}} \nonumber\\
&=&(1-\mu_{12})...(1-\mu_{k ,{k+1}})q_{m_1n_1}...q_{ {m_{k+1} n_{k+1}}}\nonumber\\
&+&\mu_{12}(1-\mu_{13})...(1-\mu_{k,{k+1}})\delta_{m_1m_2}\delta_{n_1n_2}q_{m_1n_1}\nonumber\\
&&q_{m_3n_3}...q_{m_{k+1} n_{k+1}}\nonumber\\
&+&(1-\mu_{12})\mu_{13} ...(1-\mu_{k, {k+1}})\delta_{m_1m_3}\delta_{n_1n_3}q_{m_1n_1}\nonumber\\
&&q_{m_2n_2}q_{m_4n_4}...q_{ m_{k+1} n_{k+1}}\nonumber\\
&.&\nonumber \\
&.&\nonumber \\
&.&\nonumber \\
&+&(1-\mu_{12})...(1-\mu_{k-1 ,{k+1}})\mu_{k, {k+1}}\hspace{0.5mm}\delta_{m_k m_{k+1}}\delta_{n_k n_{k+1}}\nonumber\\
&&q_{m_1n_1}q_{m_3n_3}...q_{m_{k-1}n_{k-1}}q_{m_{k+1}n_{k+1}}\nonumber\\
&+&\mu_{12} \mu_{13}(1-\mu_{14})...(1-\mu_{k, {k+1}})\delta_{m_1m_2}\delta_{n_1n_2}\nonumber\\
&&\delta_{m_1m_3}\delta_{n_1n_3}q_{m_1n_1}q_{m_4n_4}...q_{m_{k+1} n_{k+1}}\nonumber\\
&.&\nonumber \\
&.&\nonumber \\
&.&\nonumber \\
&+&\mu_{12}...\mu_{k-1, {k+1}}(1-\mu_{k,{k+1}})\delta_{m_1m_2}\delta_{n_1n_2}...\delta_{m_1m_k}\delta_{n_1n_k}\nonumber\\
&&q_{m_1n_1}q_{m_{k+1}n_{k+1}}\nonumber\\
&+&\mu_{12}...\mu_{k ,{k+1}}\hspace{0.5mm}\delta_{m_1m_2}\delta_{n_1n_2}...\delta_{m_1m_{k+1}}\delta_{n_1n_{k+1}}q_{m_1n_1}.
\label{qm1n1-mknk}
\end{eqnarray}	
Here, between every two individual channels we have defined a correlation
degree $\mu_{jl}$  with $0\leq \mu_{jl}\leq 1$  which correlates the channel
$j$ to the  channel $l$ ($j\neq l$ ).  Thus, for $k+1$ parties we have
$\frac{k(k+1)}{2}$ correlation degrees $\mu_{jl}$.
For instance,  $\mu_{12}$  correlates the channel  \emph{one} and
\emph{two},  $\mu_{k,{k+1}}$ correlates the channel   \emph{k}  and
Bob's channel, etc. If $\mu_{jl}=0$ for all $j$ and $l$,  then the
$k+1$ channels are independent or, in other words, we are in the memoryless
(or uncorrelated) case.
If $\mu_{jl}=1$ for all $j$ and $l$,  we have a \emph{fully correlated}
Pauli  channel. For other values of  $\mu_{jl}$ other than zero and
one,  the channel (\ref{k+1-channel}) is partially correlated, and it
generalises the bipartite case considered in Sect. \ref{1-3}.
We will now show examples of multipartite systems for which
$U_{\textmd{min}}^{\textmd{lo}}$  or/and $U_{\textmd{min}}^{\textmd{g}}$ are
determined.

\subsection{$k$ copies of a Bell state and a correlated Pauli channel}

The first example is  a {correlated} Pauli channel
(\ref{k+1-channel}) and  $k$ copies of the  Bell state. Noise here acts just
on the  Alices' subsystem.
Starting from the Bell state
$\ket{\psi_{00}}= \frac{1}{\sqrt d}\sum_{j=0}^{d-1}\ket{jj}$,
the set of the other maximally entangled Bell states is  denoted by
$\ket{\psi_{mn}}=(V_{mn}\otimes\id)\ket{\psi_{00}} $, for $m,n=0,1,...,d-1$.
It can be proved that the von Neumann entropy is
invariant under arbitrary unitary rotation $ U^{a_1...a_k}$ of the state
$\rho_{00}^{a_1b_1}\otimes...\otimes \rho_{00}^{a_kb_k}$ after application
of the channel  $\Lambda^{\textmd{P}}_{\textmd{a}_1...\textmd{a}_k}$
\cite{zahra-multi}. Moreover, the channel output entropy
can be written as
\begin{eqnarray}
&&S\Bigg(\Lambda^{\textmd{P}}_{\textmd{a}_1...\textmd{a}_k}\left(\left(U^{\textmd{a}_1...\textmd{a}_k}\otimes\id^{\textmd{b}_1...\textmd{b}_k} \right)
 \big(\rho^{\textmd{a}_1\textmd{b}_1}_{00} \otimes...\otimes \rho^{\textmd{a}_k\textmd{b}_k}_{00} \big)\right.\nonumber\\
&&\left. \big( {U^{\textmd{a}_1...\textmd{a}_k}}^\dagger \otimes\id^{\textmd{b}_1...\textmd{b}_k}
  \big) \right) \Bigg)
= H\left(\{q_{\{m_in_i\}}\}\right),
\label{multi-u-ivariantt}
\end{eqnarray}
where $H\left(\{ p_{i}\}\right)
=-\sum_{i} p_{i} \log p_{i}$ is the Shannon entropy. Consequently, the channel output entropy  is just determined by the channel
 probabilities $q_{\{m_in_i\}}$ and it is invariant under
 unitary encoding. Therefore, both local encoding and global encoding lead 
to the same capacity in eqs. (\ref{local-dc-covariant}) and
(\ref{global-dc-covariant}). This is given by
\begin{subequations}
 \begin{eqnarray}
\label{c-bell-k-copy}
C^{k\textmd{-copy}\textmd{}}_\textmd{un,B}&=&\log d_{1}^2+\log d_{2}^2+...+\log d_{k}^2\nonumber\\
&-&H\left(\{q_{\{m_in_i\}}\}\right),\\
&&\nonumber\\
&\neq& k \hspace{1mm} C^{\textmd{one-copy}\textmd{}}_\textmd{un,B}.
\end{eqnarray}
\end{subequations}
 The subscript \textquotedblleft \textmd{B}\textquotedblright \hspace{0.2mm} refers to a Bell state. As we can see from eq. (\ref{c-bell-k-copy}), for a {correlated} Pauli channel,  the capacity of  $k$ copies of a Bell state is not additive except when $\mu_{jl}=0$ for all $j$ and $l$, i.e. the case of an {uncorrelated} Pauli channel with $ q_{\{m_in_i\}}=q_{m_1n_1}...q_{m_kn_k}$. 
In the latter ase the capacity for $k$ copies  is $k$ times  the capacity of a single copy with dimension $d^2$ given in eq. (\ref{DCcapacityd-qubitcha}). That is 
 \begin{eqnarray}
C^{k\textmd{-copy,unco}\textmd{}}_\textmd{un,B}&=&k \left (\log d^2-H\left(\{q_{mn}\}\right) \right).
\label{c-bell-k-copy-1}
\end{eqnarray}
If $\mu_{jl}=1$ for all $j$ and $l$,  i.e. the case of a fully correlated
Pauli channel with $ q_{\{m_in_i\}}=q_{mn}$, by using eq. (\ref{c-bell-k-copy}), we have
\begin{eqnarray}
C^{k\textmd{-copy}\textmd{}}_\textmd{un,B,f}&=&\log d_{}^2+...+\log d_{}^2-H\left(\{q_{mn}\}\right)\nonumber\\
&=&k\left(\log d_{}^2- \frac {H\left(\{q_{mn}\}\right)} {k} \right).
\label{c-bell-k-copy-d}
\end{eqnarray}
Since $H\left(\{q_{mn}\}\right)$ is  constant, in the limit of many copies $k$, by using eq. (\ref{c-bell-k-copy-d}), we can reach the capacity $\log d^2$ per single copy. This is the highest capacity that we can reach for a  $d^2$ dimensional  system.

\subsection{$k$ copies of a Bell diagonal state and a fully correlated Pauli
channel \label{fully-Pauli}  }

Here, we give another example for which the capacity is exactly determined.
This is the case of  $k$
copies of a Bell diagonal state and a {fully correlated}  Pauli channel,
namely when  $\mu_{jl}=1$ for all $j$ and $l$.
For $d=2$ the channel, for  an arbitrary number of  parties, can be written as
\begin{eqnarray}
\Lambda  ^{\textmd{f}}(\xi)= \sum_m  q_m (\sigma_m \otimes...\otimes \sigma_m) \xi (\sigma_m \otimes...\otimes \sigma_m),
\label{fully-cor-channel-2}
\end{eqnarray}
where $ \sum_{m=0}^3 q_m = 1$.
The superscript   \textquotedblleft \textmd{f}\textquotedblright \hspace{0.2mm} refers to a fully correlated Pauli channel. 
As in the previous case it can be proved that
$U^{\textmd{lo}}_{\textmd{min}}=U^{\textmd{g}}_{\textmd{min}}=I$
\cite{zahra-multi}, and therefore the super dense coding capacities with both
local encoding and global encoding are
the same. According to eq. (\ref{local-dc-covariant}),
the capacity of $k$  copies of a
Bell diagonal state, when the states are sent through a {fully correlated}
Pauli channel (\ref{fully-cor-channel-2}), is additive, i.e.
\begin{eqnarray}
C^{k\textmd{-copy}}_\textmd{un,Bd,f}&=&k\big(2-S( \rho_{\textmd{Bd}})\big)\nonumber\\
&=& k \hspace{1mm}C^{\textmd{one-copy}}_\textmd{un,Bd,f} .
\label{capacity-BD}
\end{eqnarray}
The capacity (\ref{capacity-BD}) shows that for fully correlated channels no
information at all is lost to the environment and this class of channels
behaves like a noiseless one.

For k copies of a Bell state, by using eq. (\ref {capacity-BD}), and the
purity of a Bell state, we have
\begin{eqnarray}
C^{k\textmd{-copy}}_\textmd{un,B,f}&=&2k ,
\label{k-copy-bell}
\end{eqnarray}
which is the highest amount of information transfer for $2k$ parties where
each of them has  a two-level system.

\begin{table*}
  \centering
\begin{tabular}{|  c ||    c  |    c  | c l |l| }\hline
\diaghead{\theadfont Diag ColumnmnHead II}%
{ \\Resource state}{  \\ Channel }&
\thead{   {Correlated Pauli channel} \\only on the Alices' sides\\ (arbitrary dimension) }   &\thead{  Fully correlated Pauli channel  \\$  (\mu_{jl}=1)$ and ($d=2$) }  & \thead {Uncorrelated depolarising channel \\ $(\mu_{jl}=0)$ (arbitrary dimension)  } &\\
\hline\hline

{\footnotesize $\mathrm{k}$ copies of a  Bell state } & {\footnotesize eq. (\ref{c-bell-k-copy})}   &    {\footnotesize eq. (\ref{k-copy-bell})}&  {\footnotesize  eq. (\ref{2})} &  \\ \hline

 {\footnotesize $\mathrm{k}$ copies of a Bell diagonal state } &\footnotesize 
{\color{red} open } & {\footnotesize eq. (\ref{capacity-BD})} & {\footnotesize  eq. (\ref{2})}   & \\ \hline

 {\footnotesize GHZ state with 2k parties  } &  \footnotesize  {\color{red} open }  & { \footnotesize eq. (\ref{k-copy-ghz})}&   \footnotesize  {\color{red} open }  & \\ \hline 

{\footnotesize  $\mathrm{k}$ copies of an arbitrary state $\rho^{\textmd ab}$} &   \footnotesize {\color{red} open } &  \footnotesize  {\color{red} open }  &{\footnotesize  eq. (\ref{2})} & \\ \hline
\end{tabular}
\caption{A summary of the solved examples of  multipartite resource states and channels for super dense coding capacity with unitary encoding. Here, some of the unsolved examples are also mentioned as \textquotedblleft \textmd{open}\textquotedblright.}
\label{table2}
\end{table*}

\subsection{GHZ state and a fully correlated Pauli channel}

Another example for which we can determine  both unitaries
$U_{\textmd{min}}^{\textmd{lo}}$ and $U_{\textmd{min}}^{\textmd{g}}$
is a $\ket{GHZ}$ state of 2-dimensional subsystems distributed between
$2k - 1$ Alices and a single Bob. The channel here is a fully correlated
Pauli channel, as defined via eq. (\ref{fully-cor-channel-2}).
For a system of $2k$ parties, the $\ket{GHZ}$ state can be written as
\begin{eqnarray}
\ket{GHZ}_{2k}=\frac{1}{\sqrt{2}}\sum_{j=0}^1  \ket {j^{(1)}...j^{(2k)}}.
\end{eqnarray}
Since the minimum value of the von Neumann entropy is
zero, and since a $ \ket{GHZ}$ state is invariant under the action of a
fully correlated Pauli channel, we have
\begin{eqnarray}
&&S\bigg(\Lambda^{\textmd{f}}_{\textmd{a}_1...\textmd{a}_{2k-1}\textmd{b}}\Big( \ket{GHZ}_{2k}\bra{GHZ}    \Big)\bigg)\nonumber\\
    &=&S\bigg(\sum_m  q_m  (\sigma_m \otimes...\otimes \sigma_m) \Big(   \ket{GHZ}_{2k}\bra{GHZ}   \Big)\nonumber\\
&&   (\sigma_m \otimes...\otimes \sigma_m) \bigg) \nonumber\\
&&\nonumber\\
&=&S  \left(  \ket{GHZ}_{2k}\bra{GHZ} \right)=0.
\label{}
\end{eqnarray}
where in the last line, we used the fact that  $\ket{GHZ}_{2k}$ is invariant 
under unitary transformation $\sigma_m \otimes...\otimes \sigma_m$. 
Therefore, by using $U_{\textmd{min}}^{\textmd{lo}}= 
U_{\textmd{min}}^{\textmd{g}}=\id$, we have vanishing output entropy. 
Then, the super dense coding capacity,  according to eq. 
(\ref{local-dc-covariant}), reads
\begin{eqnarray}
C^{\textmd{f}}_\textmd{un,GHZ}=2k.
\label{k-copy-ghz}
\end{eqnarray}
Here, the {fully correlated}  Pauli channel, for a $\ket{GHZ}$ state, behaves like a noiseless channel and again no information is lost through the channel.

\

\subsection{$k$ copies of an arbitrary state and  an uncorrelated depolarising channel}

The last example for which we determine  the capacity   is the case of  $k$
copies  of an arbitrary state $\rho$,
each  in  dimensions $d^2$,  in the presence of an {uncorrelated} depolarising  channel $ \Lambda_{{\textmd{a}_1...\textmd{a}_k\textmd{b}_1...\textmd{b}_k}}^{\textmd{dep}}$ \cite{zahra-multi}.  This is a generalisation  of the bipartite case with the resource stat $\rho$ and the channel  $\Lambda_\textmd{ab}^{\textmd{dep}}$,
considered in Sect. \ref{two-sid}.  For this case,  the super dense coding capacity is given by \cite{zahra-multi}

\begin{eqnarray}
C^{\textmd{k-copy}}_{\textmd{un,dep}}&=&k\bigg(\log d+ S\left(\Lambda_{\textmd{b}}^\textmd{dep} \left(
\rho_{\textmd{b}}\right)\right)-S\left(\Lambda_{{\textmd{a}\textmd{b}}}^{\textmd{dep}}  \left( \rho^{\textmd{}\textmd{}}\right)  \right)\bigg),
\nonumber\\
\label{2}
\end{eqnarray}
which is $k$ times  the capacity of a single copy  given in eq. (\ref{DC-2usedepo-any-state}).

 In Table \ref{table2},  the above solved examples for the
multipartite super dense coding capacity with unitary encoding have been summarised.  Some of the  unsolved (open) examples of multipartite resource states and channels are also indicated.

\section{Conclusions \label{conc}}

In summary, we reviewed in a unified way the performance of the super dense
coding protocol in the presence of (multi-partite) covariant noisy channels,
considering both unitary and non-unitary encoding.
Regarding both types of encoding, it was shown that the problem of finding the
super dense coding capacity reduces to the easier problem of finding a
unitary operator (for unitary encoding) or CPTP map (for non-unitary encoding)
which is applied to the initial state such that it minimises
the von Neumann entropy after the action of the channel.
We then discussed in particular the case of Pauli channels, that is a broad
class of covariant noisy channels including memoryless and correlated types
of noise. We gave explicit examples of Pauli channels and initial states
for which the super dense coding capacity can be calculated analytically.
We also provided explicit examples of  non-unitary pre-processing
which can improve the super dense coding capacity in comparison to only
unitary encoding.\\
{\bf Acknowledgments}:  The research is supported by the National Research Foundation and Ministry of Education in Singapore and Deutsche Forschungsgemeinschaft (DFG) in Germany.

\end{document}